\newif\ifnotes

\documentclass[conference,twocolumn]{IEEEtran}
\IEEEoverridecommandlockouts
\usepackage{multicol}
\usepackage{algpseudocode}
\usepackage[normalem]{ulem}
\usepackage{setspace}
\usepackage[usenames,dvipsnames,table]{xcolor}
\usepackage{soul}
\ifnotes
  \usepackage[nomargin,inline,draft]{fixme}
\else
  \usepackage[nomargin,inline,final]{fixme}
\fi

\usepackage{array}
\newcolumntype{C}[1]{>{\centering\arraybackslash}p{#1}}

\usepackage[utf8]{inputenc}
\usepackage[
  sorting=none,
  style=numeric-comp,
  backend=biber,
  isbn=false,
  url=true,
  doi=false,
  url=false,
  mincrossrefs=100,
  maxnames=5,
  firstinits=true
]{biblatex}

\AtEveryBibitem{%
  \clearfield{note}%
}

\usepackage[T1]{fontenc}
\usepackage{microtype}
\usepackage{pstricks}
\usepackage{graphicx}
\usepackage{amsmath}
\usepackage{amssymb}
\usepackage{xspace}
\usepackage{listings}
\usepackage{multirow} 
\usepackage[most]{tcolorbox}
\usepackage{booktabs}
\usepackage{tabularx}
\usepackage{xspace}
\usepackage{multirow}
\usepackage{pifont}
\usepackage{amsthm}

\usepackage{mathtools}

\usepackage{relsize}
\usepackage{flushend}

\usepackage{caption}
\usepackage{subcaption}
\usepackage{mdframed}

\definecolor{highlight1}{HTML}{F58F33} 
\definecolor{highlight2}{HTML}{2F7BA3} 

\colorlet{highlight1_transparent}{highlight1!50} 
\colorlet{highlight2_transparent}{highlight2!50} 

\newcommand{\commentout}[1]{}

\usepackage{hhline}

\AtEveryBibitem{\clearname{editor}}

\definecolor{white}{RGB}{255,255,255}
\definecolor{vrpink}{RGB}{255,0,127}
\definecolor{vrblue}{RGB}{30,144,255}
\definecolor{vrolive}{RGB}{85,107,47}
\definecolor{vrroyalblue}{RGB}{65,105,225}
\definecolor{brgreen}{RGB}{100,200,70}
\definecolor{ivsalmon}{RGB}{255,160,122}
\definecolor{vrlpink}{RGB}{255,192,203}
\definecolor{mvcol}{RGB}{5,150,25}

\usepackage[obeyFinal]{todonotes}
\usepackage{soul}

\FXRegisterAuthor{vv}{avr}{\color{vrpink}[vince]}
\FXRegisterAuthor{rvv}{arvr}{\color{red}[vince]}

\newcommand{\jd}[1]{\ifnotes{\textbf{\textcolor{blue}{JD: #1}}}\fi}

\xdefinecolor{DarkGreen}{cmyk}{0.8,0,0.8,0.3}

\fxusetheme{color}
\fxuseenvlayout{color}

\usepackage{graphicx}
\graphicspath{{../figures/}}
\DeclareGraphicsExtensions{.pdf,.jpeg,.png,.jpg}

\usepackage{comment}

\usepackage{algorithm}
\usepackage{algpseudocode}

\algblockdefx{As}{EndAs}[1]%
{\textbf{as a }#1}%
{\textbf{}}
\algblockdefx{All}{EndAll}[1]%
{\textbf{all replicas}#1}%
{\textbf{}}
\algblockdefx{Timeout}{EndTimeout}[1]%
{\textbf{upon timeout}#1}%
{\textbf{}}
\algblockdefx{Func}{EndFunc}[1]%
{\textbf{function }#1}%
{\textbf{}}
\algblockdefx{WFunc}{EndWFunc}[3]%
{\textbf{function }#1\textbf{ where }#2\textbf{ is }#3}%
{\textbf{}}
\algblockdefx{WFuncx}{EndWFuncx}[3]%
{\begin{tabular}{r@{\hspace{0.03in}}l}\textbf{function } & #1\\\textbf{ where } & #2\textbf{ is }#3\end{tabular}}%
{\textbf{}}
\algblockdefx{WFuncb}{EndWFuncb}[3]%
{\textbf{function }#1\textbf{ where }#2\textbf{ is }#3}%
{\textbf{}}
\algtext*{EndWFuncb}%
\algblockdefx{WWFunc}{EndWWFunc}[5]%
{\begin{tabular}{l@{\hspace{0.03in}}r@{\hspace{0.03in}}l}\textbf{function }#1&\textbf{where}&#2\textbf{ is }#3\\&\textbf{and}&#4\textbf{ is }#5\end{tabular}}%
{\textbf{}}
\algblockdefx{WWWFunc}{EndWWWFunc}[7]%
{\begin{tabular}{l@{\hspace{0.03in}}r@{\hspace{0.03in}}l}\textbf{function }#1&\textbf{where}&#2\textbf{ is }#3\\&\textbf{and}&#4\textbf{ is }#5\\&\textbf{and}&#6\textbf{ is }#7\end{tabular}}%
{\textbf{}}
\algblockdefx{Funcs}{EndFuncs}[1]%
{\textbf{functions }#1}%
{\textbf{}}
\algblockdefx{Funcsb}{EndFuncsb}[1]%
{\textbf{functions }#1}%
{\textbf{}}
\algtext*{EndFuncsb}%
\algblockdefx{Funcb}{EndFuncb}[1]%
{\textbf{function }#1}%
{\textbf{}}
\algtext*{EndFuncb}%
\algnewcommand\lFor[2]{\State\textbf{for}\ #1\ \textbf{do}\ #2}
\algnewcommand\lIf[2]{\State\textbf{if}\ #1\ \textbf{then}\ #2}

\algnewcommand\algorithmicforeach{\textbf{for each}}
\algdef{S}[FOR]{ForEach}[1]{\algorithmicforeach\ #1\ \algorithmicdo}

\newcommand{\hide}[1]{}
\newcommand{\nemo}{\texttt{NEMO}} 
\newcommand{\nemonopq}{\texttt{NEMO\_no\_pq}} 



\usepackage{mathtools}
\newtagform{blue}{\color{BlueViolet}(}{)}

\begin{document}

\def\sectionautorefname{Sec.}
\def\subsectionautorefname{Sec.}
\def\figureautorefname{Fig.}
\def\tableautorefname{Tab.}
\def\algorithmautorefname{Alg.}
\def\equationautorefname{Eq.}

\title{\nemo{}: Faster Parallel Execution for Highly Contended Blockchain Workloads}

\author{
    \IEEEauthorblockN{François Ezard\IEEEauthorrefmark{1}, Can Umut Ileri\IEEEauthorrefmark{2}, J\'er\'emie Decouchant\IEEEauthorrefmark{1}}
    \IEEEauthorblockA{
        \IEEEauthorrefmark{1}Delft University of Technology, \IEEEauthorrefmark{2}IOTA Foundation \\
        francois.ezard@gmail.com, canumut.ileri@iota.org, j.decouchant@tudelft.nl
    }
}

\maketitle

\begin{abstract}
Following the design of more efficient blockchain consensus algorithms, the execution layer has emerged as the new performance bottleneck of blockchains, especially under high contention. 
Current parallel execution frameworks either rely on optimistic concurrency control (OCC) or on pessimistic concurrency control (PCC), both of which see their performance decrease when workloads are highly contended, albeit for different reasons.  In this work, we present \nemo{}, a new blockchain execution engine that combines OCC with the object data model to address this challenge. \nemo{} introduces four core innovations: (i) a greedy commit rule for transactions using only owned objects; (ii) refined handling of dependencies to reduce re-executions; (iii) the use of incomplete but statically derivable read/write hints to guide execution; and (iv) a priority-based scheduler that favors transactions that unblock others. Through simulated execution experiments, we demonstrate that \nemo{} significantly reduces redundant computation and achieves higher throughput than representative approaches.  For example, with 16 workers \nemo{}'s throughput is up to 42\% higher than the one of Block-STM, the state-of-the-art OCC approach, and 61\% higher than the pessimistic concurrency control baseline used.
\end{abstract}


\section{Introduction}

Blockchain technologies have been rapidly evolving since Bitcoin~\cite{BitcoinWhitepaper}, which demonstrated that financial services can be implemented in a decentralized manner~\cite{Dipetrans, Concerto}. Another key milestone was the introduction of the Ethereum Virtual Machine (EVM)~\cite{Ethereum}, which allows transactions to invoke code specified as a \textit{smart contract}~\cite{Forerunner, MBPS}. Smart contracts extended the range of possible uses of blockchain technologies~\cite{MeteringTheMeter}. 
However, while the potential of blockchains has been unfolding,  they have also been facing adoption issues. In the eye of the public, one of the main limitation of blockchain technologies is their lower performance compared to traditional services~\cite{MeteringTheMeter, Stingray}. For example, Bitcoin and Ethereum can, respectively, only commit 7 and 30 transactions per second (TPS). In comparison, Visa can process up to 65,000 per second~\cite{MeteringTheMeter}. In addition, traditional blockchains like Bitcoin and Ethereum execute transactions sequentially to ensure that all validators end up in the same state, which significantly limits their execution throughput, as it fails to efficiently utilize modern multicore CPU architectures~\cite{DMVCC, Stingray}.

Since consensus was the primary performance bottleneck in early blockchains such as Bitcoin and Ethereum\cite{Chiron}, efficient transaction execution was not a critical concern, and coupling transaction ordering with consensus was a reasonable design choice. However thanks to recent approaches that improve the performance of consensus algorithms~\cite{Narwhal, Bullshark, HotStuff, Mysticeti}, the performance bottleneck has been shifted to transaction execution~\cite{Chiron, DMVCC, OCCDA, Anthemius, Block-X, Batch-Schedule-Execute}. 
As a consequence, modern blockchains~\cite{AptosWhitepaper,SuiWhitepaper} have adopted a modular software architecture where consensus is decoupled from execution following an \textit{Order-Execute} model~\cite{Thunderbolt}. 
These \textit{lazy} blockchains~\cite{LazyBlockchainClient} decouple components for independent optimization. 
Parallel execution improves execution throughput while still ensuring that all validators arrive at a consistent final state. More formally, this requirement is called \textit{deterministic serializability}: the results of parallel execution should be identical to sequentially executing all transactions according to the order defined by the consensus layer~\cite{OptME, DMVCC, Batch-Schedule-Execute}. Data accesses in blockchain workloads are highly skewed \cite{ParallelEVM} leading to highly contended workloads, which makes it challenging to maintain high performance whilst ensuring deterministic serializability.

\begin{table*}[h]
     \centering
    \caption{Comparison of \nemo{} with the related work.}
    \label{paper:tab:comparison-related-work}
    \begin{tabular}{|C{3.0cm}|C{1.6cm}|c|c|C{2.3cm}|}
    \hline
    & \textit{Transaction \newline Model} & \textit{Application} & \textit{PCC/OCC} & \textit{Efficient under \newline high contention} \\ \hline
    DOCC~\cite{DOCC}, Gria~\cite{Gria}, Dodo~\cite{Dodo} & - & \cellcolor{red!25}Deterministic Database & \cellcolor{green!25}OCC & \cellcolor{green!25}\ding{52} \\ \hline
    ParBlockchain~\cite{ParBlockchain}, PEEP~\cite{PEEP} & \cellcolor{red!25}Account & \cellcolor{green!25}Blockchain Execution & \cellcolor{red!25}PCC & \cellcolor{red!25}\ding{54} \\ \hline
    Block-STM~\cite{Block-STM}, RapidLane~\cite{RapidLane} & \cellcolor{yellow!25}Resource & \cellcolor{green!25}Blockchain Execution & \cellcolor{green!25}OCC & \cellcolor{red!25}\ding{54} \\ \hline
    Chiron~\cite{Chiron} & \cellcolor{yellow!25}Resource & \cellcolor{red!25}Straggler Execution & \cellcolor{green!25}OCC & \cellcolor{green!25}\ding{52} \\ \hline
    Shardines~\cite{Shardines} & \cellcolor{yellow!25}Resource & \cellcolor{green!25}Blockchain Execution & \cellcolor{yellow!25}Hybrid & \cellcolor{red!25}\ding{54} \\ \hline
    Pilotfish~\cite{Pilotfish}, Sui~\cite{SuiLutris} & \cellcolor{green!25}Object & \cellcolor{green!25}Blockchain Execution & \cellcolor{red!25}PCC & \cellcolor{red!25}\ding{54} \\ \hline
    ParallelEVM~\cite{ParallelEVM} & \cellcolor{red!25}Account & \cellcolor{green!25}Blockchain Execution & \cellcolor{yellow!25}Hybrid & \cellcolor{green!25}\ding{52} \\ \hhline{|=|=|=|=|=|}
    \nemo{} (this work) & \cellcolor{green!25}Object & \cellcolor{green!25}Blockchain Execution & \cellcolor{green!25}OCC & \cellcolor{green!25}\ding{52} \\ \hline
    \end{tabular}
    \vspace{0.2cm}

    {\small \textbf{Abbreviations:} PCC = Pessimistic Concurrency Control; OCC = Optimistic Concurrency Control.}
\end{table*}

Concurrency control manages conflicts that might arise when executing transactions in parallel. There are two main concurrency control approaches: \textit{pessimistic} and \textit{optimistic}. Pessimistic concurrency control (PCC)~\cite{PEEP,SuiWhitepaper}, also called static parallelism, executes possibly conflicting transactions sequentially, which requires prior knowledge of the transactions' read and write sets, and often overestimates them, to identify and avoid potential conflicts. Optimistic concurrency control (OCC)~\cite{Block-STM,RapidLane,Chiron}, also called dynamic parallelism, does not require prior knowledge of the read and write sets. Instead, it optimistically assumes that all transactions are independent and conflicts are detected in the validation step, which may trigger a re-execution of transactions. Both PCC and OCC experience a performance degradation under high contention, though for different reasons. PCC incurs significant overhead due to increased fallback to sequential execution, while OCC experiences frequent transaction re-executions caused by conflicts. Given that blockchain workloads are often highly contended~\cite{Anthemius,DMVCC,OCCDA, ParallelEVM}, preserving the performance gains of parallel execution in such environments is critically important.  

Besides concurrency control approach, another important factor that influences the frequency of conflicts in blockchain systems is the underlying data model. Bitcoin employs the \textit{unspent transaction output} (UTXO) model, where coins are represented as outputs of previous transactions, and transactions themselves are the primary means of storing and transferring value. Ethereum uses the \textit{account} data model, in which each smart contract and user has a persistent account address~\cite{Chartalist}. In this model, the state of each account is maintained and can be modified by transactions~\cite{UTXOvsAccount}. However, because transactions in this model do not encapsulate the complete end state, the likelihood of conflicts is increased. This limits opportunities for parallel execution and creates a potential performance bottleneck. To address these limitations, newer blockchains have introduced alternative data models to improve parallelism, reduce contention and improve scalability.  For instance, Block-STM, Aptos's execution engine~\cite{AptosWhitepaper}, uses OCC with the \textit{resource}-based data model built on top of the Move virtual machine (VM)~\cite{MoveVm}, where resources are associated to specific accounts. Based on the same VM, Sui~\cite{SuiWhitepaper}, and IOTA~\cite{iota_rebased_2024} which adopt the same protocol, take a different approach with PCC and an \textit{object}-based data model, treating each object as an independent unit of state  with a unique global address.

In this work, we design \nemo{}, a parallel blockchain execution engine designed to sustain high throughput even under highly contended workloads. \nemo{} is the first system to combine optimistic concurrency control with the object-based data model, leveraging the benefits of both paradigms to maximize parallelism and reduce execution overhead under high contention. \nemo{} does not require prior knowledge of the read/write sets, but is able to leverage partial prior knowledge to dynamically schedule the execution of transactions. \nemo{} handles potential conflicts following the OCC paradigm.   
To achieve this, \nemo{} introduces four key innovations: 
\begin{enumerate}
    \item A greedy commit rule for transactions operating solely on owned objects, enabling a fast-path to execute without coordination; 
    \item A refined dependency tracking mechanism to minimize transaction re-executions;
    \item The use of incomplete but statically derivable read/write hints to guide dynamic scheduling; 
    \item A priority-based scheduler that favors transactions which unblock dependent ones. 
\end{enumerate} 

These techniques allow \nemo{} to deliver consistent performance improvements, especially in scenarios where traditional OCC and PCC systems would degrade due to conflict management overhead. 
In summary, this paper makes the following \textbf{contributions}:
\begin{itemize}
    \item We adapt the state-of-the-art OCC execution engine, Block-STM~\cite{Block-STM}, to support the object data model, introducing a novel fast-path via a greedy commit rule for transactions that use only owned objects.
    \item We propose new techniques that specifically target performance under high-contention workloads.
    \item We provide a comprehensive evaluation using realistic blockchain workloads, demonstrating that \nemo{} achieves significant performance gains over existing systems.  
\end{itemize}

\section{Related Work}

\begin{figure*}[t]
    \centering
    \includegraphics[width=.80\textwidth]{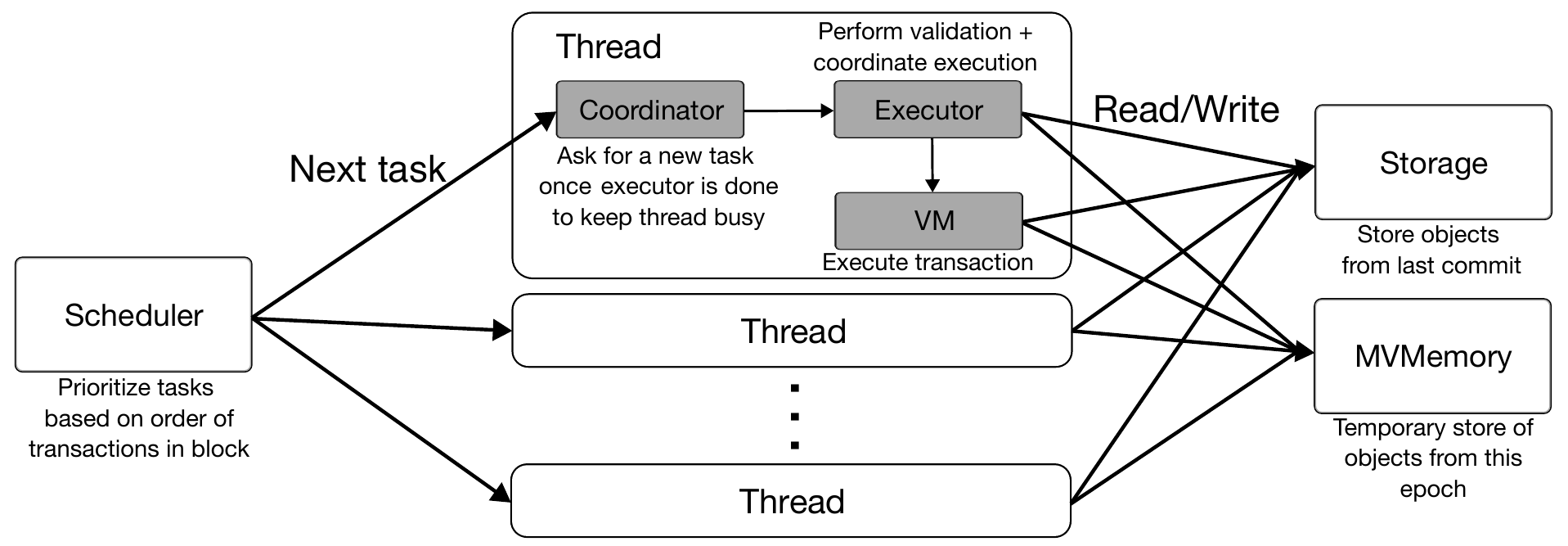}
    \caption{Block-STM's architecture}
    \label{paper:fig:block-stm-architecture}
\end{figure*}

Block-STM~\cite{Block-STM} is a parallel execution engine for smart contracts that uses OCC  and is currently used in production in the Aptos~\cite{AptosWhitepaper} blockchain. Block-STM optimistically assumes that all transactions are independent and executes them in parallel recording their read/write sets. Those sets are then used to validate the results of execution, which leads to re-execution if a transaction fails validation. Once all transactions have been validated and there are no more tasks to execute, Block-STM lazily commits the block of transactions. This results in the same state as having sequentially executed the block. Block-STM is able to achieve great performance for low contention workloads as it is able to parallelize the workload efficiently, but it struggles under high contention due to frequent re-executions.

RapidLane~\cite{RapidLane} builds on top of Block-STM and introduces so called \textit{deferred objects} to eliminate read-write conflicts for frequently mutated objects. It does by using a heuristic to avoid reading the actual value, and validates the decision taken during the commit phase. There are two main problems with this approach the first one being that the performance heavily relies on the heuristic. The second issue is that RapidLane restricts the types that can be wrapped into a deferred object to integer counters and small-sized strings (up to 256 bytes), which greatly limits the applicability of deferred objects.

Chiron~\cite{Chiron} is a protocol that builds on top of Block-STM to help straggler nodes catch up. It uses prior knowledge of the read/write sets extracted from the non-straggling nodes to build a dependency graph. This graph is then used to perform guided parallel execution, where transactions are only executed if they do not have any unresolved dependencies and blocking transactions are prioritized.

PEEP~\cite{PEEP} is a blockchain execution engine that uses PCC, thus requiring prior knowledge of the transactions' read/write sets which it assumes can be obtained either via execution simulation or static analysis. Transactions then need to acquire all the locks needed before they can be executed by a worker thread. PEEP struggles under high contention workloads as it limits the amount of parallelism and suffers from the overhead introduced by locking.

Sui~\cite{SuiLutris}'s execution engine assigns version numbers to all the objects to enforce a causal order between transactions. Once all object versions a transaction takes in are available, the transaction is ready to be executed and increments the versions of all objects it writes. This approach requires complete prior knowledge of the transactions' read/write sets, which is done by leveraging the object model to obtain the exhaustive set of objects that a transaction could read/write. However, objects in this set are not necessarily used during execution as the transaction logic depends on the global state at the time of execution. This results in an overly cautious approach that fails to take advantage of the limited parallelism opportunities in high contention workloads.

ParallelEVM~\cite{ParallelEVM} uses a novel \textit{operation-level} concurrency control algorithm, which handles conflicts at the operation level rather than at the transaction level. This increases the amount of work that can be done in parallel which has a greater impact for high contention workloads. ParallelEVM was developed for Ethereum and relies on the assumption that conflicts affect only a few operations. It is unclear whether this assumption holds for other blockchains and for the object data model.

Some approaches~\cite{Shardines,Pilotfish,ParBlockchain} have looked into sharded execution as a way to split the load across several machines. These approaches introduce some added complexity, and require accurate prior knowledge of the read/write sets in order to properly coordinate workers. Furthermore, this coordination is expensive as it requires cross-shard communication which is bound to frequently happen in highly contended workloads.

Related to blockchains, deterministic databases have very different workloads but also require deterministic serializability, and often follow the \textit{Order-Execute} paradigm. Recent works~\cite{DOCC,Gria,Dodo} have shown that OCC can be used even under high contention and ensure high throughput. 

Table~\ref{paper:tab:comparison-related-work} summarizes how \nemo{} differs from the state-of-the-art execution approaches.

\section{Background}

In this section, we provide a detailed description of two representative execution layers, namely Block-STM~\cite{Block-STM} and Sui~\cite{SuiWhitepaper}'s, which respectively follow the OCC and PCC approaches. We also explain why \nemo{} builds on top of Block-STM to optimize for high contention scenarios.

\textbf{Block-STM's OCC.} Figure~\ref{paper:fig:block-stm-architecture} illustrates Block-STM's architecture. The scheduler distributes tasks to execution threads. It prioritizes tasks following the order of transactions defined by a block, whilst also trying to keep all the threads busy at all times. There are two types of tasks a thread can be given, a transaction execution task or a transaction validation task, neither of which can be aborted. Transaction execution is done first and creates a validation task to verify the execution result. For an execution task a thread tries to read all input objects from the Multi-Version Memory (MVMemory) store, but ends up reading the last committed version from storage if there is no entry for that object. All writes are done to MVMemory and the final value of an object is committed to storage at the end of an epoch, i.e., when a block has been processed. For validation tasks a thread reads the latest available version of each input object, like for an execution task, and then compares their version number with the version number that was read when the transaction was executed ensuring that the resulting state is correct. Crucially, MVMemory is shared across all threads so that when a transaction fails validation, it marks all entries that it wrote as \texttt{ESTIMATE} to indicate that other transactions should not read their value as they are likely to be overwritten in a near future. The scheduler also tries to keep track of dependencies to avoid known conflicts that would likely lead to additional re-executions.

\begin{figure}[t]
    \centering
    \includegraphics[width=.95\columnwidth]{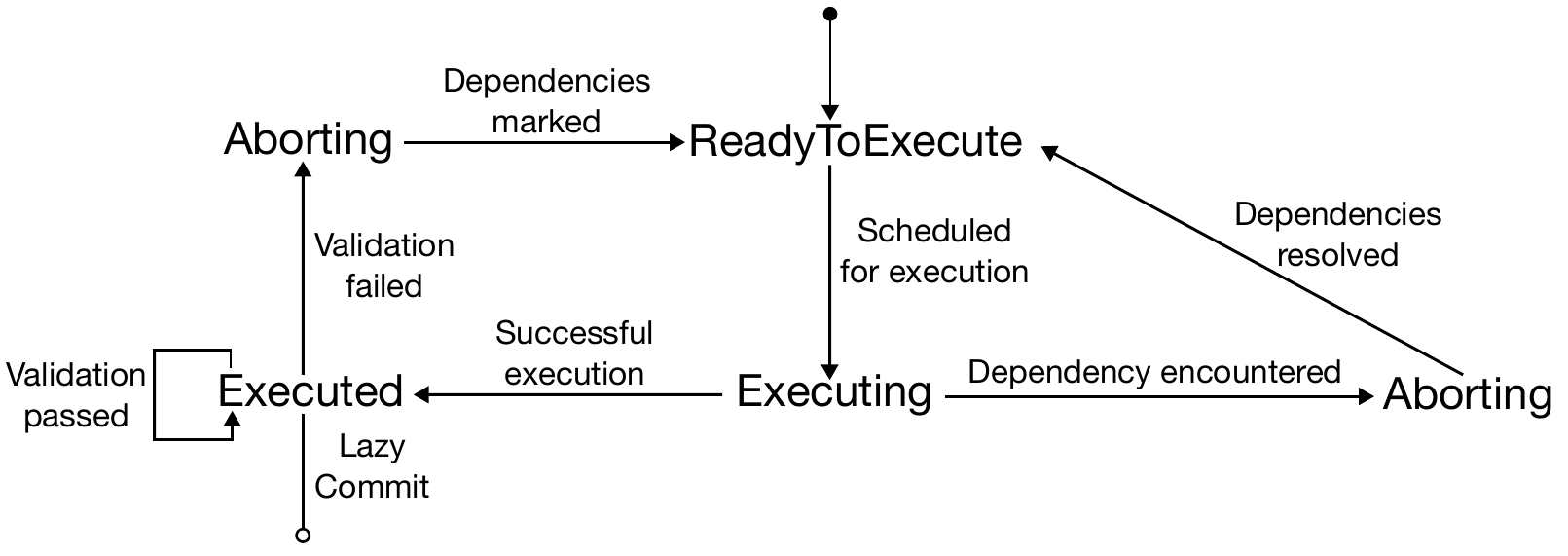}
    \caption{Transaction states and transitions between them in Block-STM}
    \label{paper:fig:block-stm-cycle}
\end{figure}

Figure~\ref{paper:fig:block-stm-cycle} shows all possible transaction states in Block-STM. Block-STM optimistically assumes that all transactions are independent, starting all of them in the \texttt{ReadyToExecute} state. The scheduler then prioritizes tasks based on the order defined by the block, and eventually every transaction will be scheduled for execution and move to the \texttt{Executing} stage. Two things can happen there, it will either encounter an entry marked as \texttt{ESTIMATE} indicating a dependency in which case the transaction is aborted, or execution will successfully complete. In the first case, the transaction goes into the \texttt{Aborting} state and remains in that state until the blocking transaction has been re-executed. Once that happens the transaction goes back into the \texttt{ReadyToExecute} state as the dependency has been resolved. In the second case, the transaction goes into the \texttt{Executed} state and awaits validation. If that validation fails then the transaction is aborted and all of its write entries are marked with the \texttt{ESTIMATE} flag to stop other transactions from reading those entries. This way of reaching the \texttt{Aborting} state is different from the previous one, as a transaction only briefly passes through it to mark dependencies. A transaction failing validation also means that all transactions that appear after it in a block and have already been validated need to be re-validated. Under high contention this can result in cascading failures, where one failed validation leads to a long a chain of other failed validations, drastically reducing throughput. An epoch concludes once all transactions are in the \texttt{Executed} state, lazily committing the entire block.

\textbf{Sui's PCC.} Sui~\cite{SuiWhitepaper} is a decentralized smart contract platform, which focuses on low-latency and is maintained by a permissionless set of validators~\cite{SuiLutris}. It follows the \textit{Order-Execute} paradigm and uses a modular approach that decouples execution from consensus~\cite{Anthemius}. A key feature of Sui is its object-centric data model in which the global state is represented by the state of all objects. There are 3 types of objects: i) \textit{immutable objects} can be read by any transactions but cannot be modified; ii) \textit{owned objects} that have a single owner which grants permission to use the object; and iii) \textit{shared objects} that do not have an owner and can be included in anybody's transactions~\cite{SuiLutris}. Critically, while smart contracts can create and mutate objects they do not store them~\cite{AllAboutObjects}, which enables more parallelism as transactions touching different object can safely be executed in parallel. Furthermore, the distinction between \textit{shared} and \textit{owned} objects allows for minimizing conflicts between transactions, as an \textit{owned} object can only be used by one transaction per epoch thus removing all possible conflicts. Another key advantage of the object data model, is that each object has its own global address, which allows Sui to provide an exhaustive, but often overestimated, set of objects a transaction may use prior to execution.

\begin{figure}[t]
    \centering
    \includegraphics[width=\columnwidth]{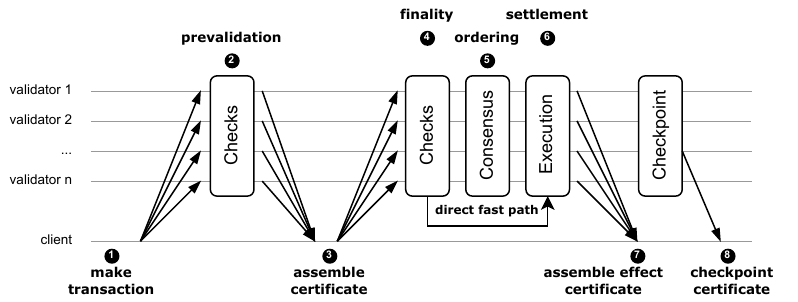}
    \caption{Transaction lifecycle (from~\cite{SuiLutris})}
    \label{paper:fig:transaction-lifecycle}
\end{figure}

The overall lifecycle of a transaction in the Sui blockchain is shown in  Figure~\ref{paper:fig:transaction-lifecycle}. Sui assumes that it has complete knowledge of the read/write sets of transactions thanks to the object model and uses this information to avoid conflicts during execution. Execution can actually be broken down into 3 different components: i) Sequencer; ii) Execution Manager; and iii) Execution Driver. The sequencer takes the output of consensus as input, and for each transaction it assigns a version number to each shared object on which the transaction operates~\cite{SuiLutris}. This effectively enforces a partial order between transactions, as a transaction is only ready to be executed once all its shared objects have the right version number. The execution manager is tasked with keeping track of the different object versions and of the status of the different transactions. Once a transaction is ready to be executed, the execution manager sends it to the execution driver, which executes the transaction using a virtual machine (VM) that is built on top of MoveVM~\cite{MoveVm}. The execution of a transaction potentially updates the version numbers of some of its shared objects, which the execution manager is made aware of. A limitation of this approach is that whilst the object model enables extracting exhaustive read/write sets ahead of time, these sets are a superset of the actual sets used during execution. This is because the actual read/write sets used during execution depend on the state at the time of execution, which is impossible to predict. In general PCC has to overestimate the read/write sets of transactions in order to prevent conflicts. In high contention workloads, PCC then typically limits parallelism and therefore has a lower throughput than OCC.  

\section{\nemo{}}

This paper introduces \nemo{} a novel blockchain execution framework that is optimized for high contention workloads. This task is challenging as improving the throughput of the execution layer is dependent on the ability to parallelize the workload and high contention workloads have limited inherent parallelism. As a result, our protocol has to be able to fully exploit the limited parallelism available.

    \subsection{Combining the object-model and OCC}

\nemo{} uses the OCC paradigm as it has been shown that the performance obtained with a lock-free approach is higher than the one obtained with a lock-based approach under scenarios with hotspots~\cite{Arete}, which are common for blockchains~\cite{ParallelEVM}.
\nemo{} then adopts the object model as it provides the greatest potential for parallelism and provides more prior knowledge about the read/write sets. 
\nemo{} combines the object model with OCC to take advantage of every opportunity to parallelize the workload whilst avoiding cascading aborts. 
\nemo{} is implemented on top of Block-STM. We believe that \nemo{} is the first protocol to combine OCC with Sui's object data model and it is also among the first protocols to optimize blockchain execution for the high contention scenarios.

\subsection{Greedy commit rule}

We have adapted Block-STM~\cite{Block-STM} to the object data model to take advantage of the parallelism it offers. This involved tweaking \texttt{MVMemory}, \texttt{Storage} and the \texttt{VM} to support objects. It is then possible to  take advantage of \textit{owned objects}, as an owned object can only be used by one transaction per epoch due to transaction certification. This means that a transaction that only uses owned objects will be independent of all other transactions in that epoch, and can safely be executed at any time. Such a transaction can also skip the validation step and be greedily committed immediately after execution. This \textit{greedy commit rule} offers a true fast-path for transactions that do not interact with the shared global state, and builds upon the existing fast-path that allows such transactions to bypass consensus. \jd{Sui Lutris only?}

\subsection{Limiting re-executions}

Transaction execution is the costliest operation in terms of time, so limiting the number of times a transaction is re-executed is crucial for performance. This is further exacerbated by the fact that tasks cannot be aborted, and so they can occupy a thread for a significant portion of time. The key to obtaining good performance when using OCC is to avoid prematurely triggering re-execution which likely will result in further re-execution.

DOCC~\cite{DOCC} only validates a transaction once all previous transactions it has a write conflict with have been committed. This enforces that every transaction can be executed at most twice, which prevents highly conflicting transactions from re-executing several times. This rule does have the downside of leaving re-execution to the last moment, which can lead to a situation where a lot of transactions are stuck waiting on the re-execution. Another issue of using this approach is that it delays finding out that a transaction will require re-execution, and uses validation as a proxy to prevent premature re-execution. \nemo{} adapts this logic to validate early, similarly to Block-STM, but to have transactions enter the \textit{Waiting} phase if they have some unresolved dependencies. For this to work, \nemo{} has to extract more information about dependencies from execution. So whereas Block-STM is only able to detect dependencies if execution read in a \texttt{ESTIMATE} marker, \nemo{} looks to extract information even when execution succeeds. This information can then be used in case the transaction fails validation later on. This approach has a higher impact on performance in high contention workloads, as there are then more dependencies between transactions which results in more frequent re-executions.

Another optimization that \nemo{} uses to limit re-executions consists in only resolving dependencies once the blocking transaction passes validation, unlike Block-STM that resolves dependencies after a successful execution. This means that transactions are blocked for longer, but also that once a dependency is resolved it is less likely to cause issues later on. This avoids having long chains of dependencies that are re-executed several times and cause multiple cascading aborts.

\subsection{Incomplete hints}

Block-STM~\cite{Block-STM} assumes zero prior knowledge about the read/write sets of all transactions and that all transactions are independent. Whilst this is useful in making Block-STM generalizable and allows it to exploit parallelism in low-contention workloads, it is not a very realistic assumption and results in an overly optimistic protocol. Under high contention this results in a lot of failing validations and thus a lot of re-executions. As mentioned earlier, one of the main advantages of our use of the object model is that it makes it easier to identify the resources that a transaction uses~\cite{SuiLutris}, since each object has its own globally unique identifier and transactions refer to shared objects explicitly. This allows one to statically extract the exhaustive set of objects a transaction may access. However, this exhaustive set is often overly pessimistic, as it includes all possible objects a transaction might touch depending on the execution path and global state, which limits parallelism. This case is especially problematic in high-contention scenarios where opportunities for parallelism are already scarce.

\nemo{} seeks a middle ground between these two extremes by leveraging partial information about the read/write sets of transactions to limit unnecessary re-executions, whilst remaining optimistic enough to exploit parallelism.
Unlike Sui however, this information does not have to be complete and \nemo{} is still correct if no prior knowledge is available. 
Ideally, the partial information should only include objects that are guaranteed to be accessed during execution, that is, objects whose usage does not depend on the runtime global state.
\nemo{} assumes that this partial information can be obtained statically by leveraging the object data model, as it exposes unique identifiers and explicit references in transaction logic. If no such information is available, \nemo{} falls back to the same assumptions as Block-STM.

\begin{figure}[t]
    \centering
    \includegraphics[width=0.6\columnwidth]{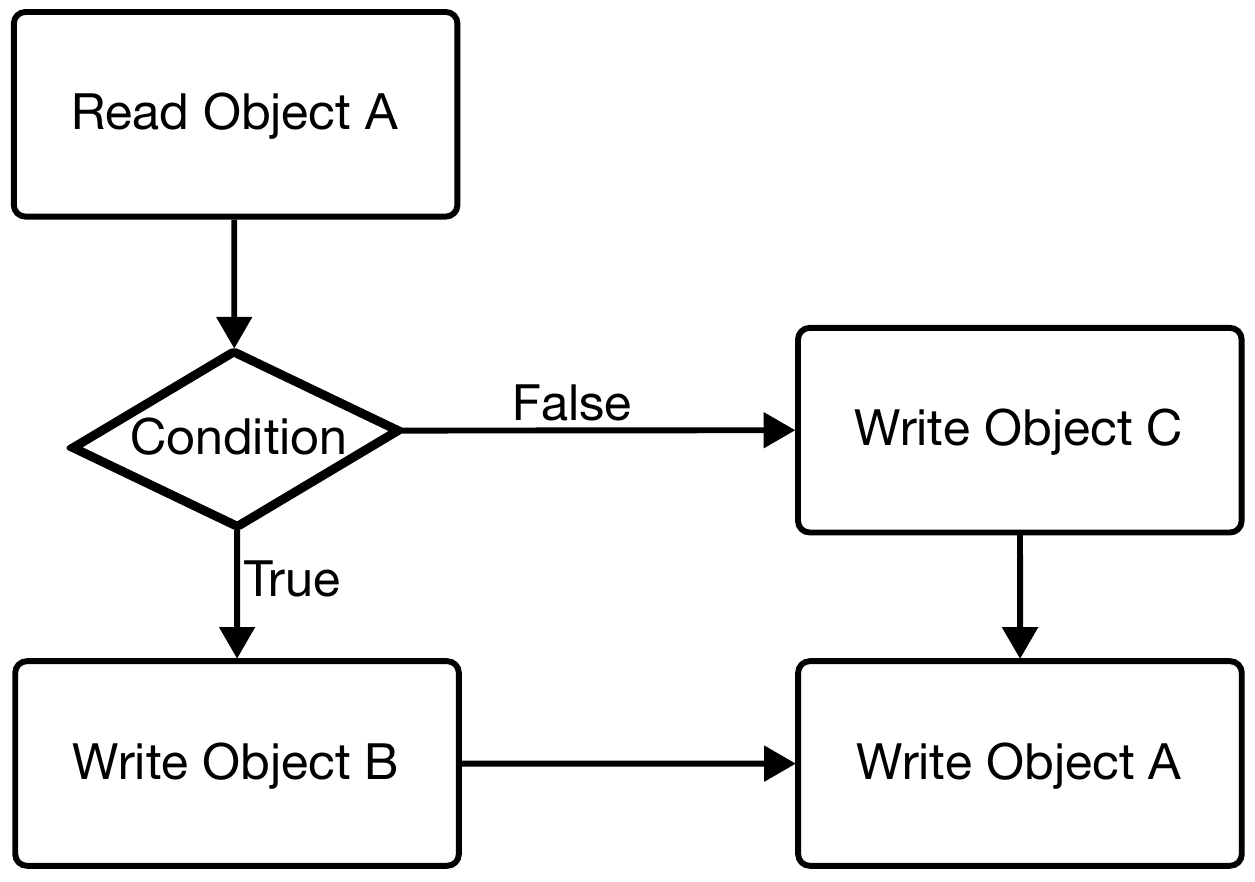}
    \caption{Example transaction logic showing how prior knowledge differs: exhaustive knowledge includes all possible writes ($A$, $B$, $C$), while partial knowledge only includes guaranteed writes ($A$).}
    \label{fig:smart-contract-logic}
\end{figure}

Figure~\ref{fig:smart-contract-logic} illustrates the difference between exhaustive and partial prior knowledge on a simple transaction example. In this example, the condition is based on the value of object~$A$ and determines whether the transaction writes to object~$B$ or~$C$. Sui would extract the exhaustive sets $read = {A}$ and $write = {A, B, C}$ to avoid all possible conflicts. In contrast, \nemo{} only includes the guaranteed accesses, i.e., $read = {A}$ and $write = {A}$, which improves parallelism by not over-constraining the schedule.

\nemo{} then uses this information to schedule conflicting transactions in order and avoid unnecessary re-executions.
During preprocessing, \nemo{} marks the known write entries and detects dependencies by checking for reads on those marked objects.
Because this prior knowledge is not exhaustive it means that it does not restrict \nemo{} from fully exploiting the available parallelism of the workload but it also means that re-execution can still happen. \nemo{} makes this tradeoff because we believe that when optimizing for high contention it is better to remain optimistic in order to take advantage of the limited opportunities to parallelize the workload.

\subsection{Dependency-based priority scheduling}

Block-STM~\cite{Block-STM} prioritizes tasks purely based on the transaction index, which means that it prioritizes tasks based on the total order defined by the block. This logic is simple and prioritizes earlier transactions to resolve dependencies through execution. The issue with this approach is that it fails to use all the available information, as it does not use the knowledge it has about dependencies obtained through previous execution. Using this approach a transaction that is independent of all other transactions but has a smaller index will be prioritized over a transaction with a larger index but on which other transactions depends.

Chiron~\cite{Chiron} showed how guided execution built on top of Block-STM could work and managed to achieve up to a 30\% speedup. Chiron was designed to help straggler nodes catch up with the help of hints obtained from validators that are ahead. It then uses these hints to perform guided execution prioritizing transactions that are blocking others. If one transaction fails validation, then Chiron falls back to Block-STM for the remainder of the block. This means that in order for Chiron to achieve better performance than Block-STM, it requires accurate and complete hints to avoid any validations failing. 

\nemo{} takes inspiration from Chiron's approach, but never falls back upon Block-STM and also uses priority scheduling for both execution and validation tasks. The scheduler maintains a priority queue of tasks where the score associated with a task is the number of transactions that directly depend on it. Tasks are then prioritized based on decreasing score. In case of a tie, the task with the smallest index has higher priority. The scheduler also maintains a set of the tasks currently in the queue in order to avoid inserting a task that is already in the priority queue. This scheduling approach prioritizes blocking transactions, which is in expectation resolves dependencies earlier resulting in a more parallelizable remaining workload. This combines well with incomplete hints, as they provides more information to the scheduler. It is worth noting that if there is no prior knowledge available then all transactions initially have a score of 0, and so the scheduler will behave in the same manner as Block-STM until dependencies are uncovered during execution.

\section{Performance Evaluation}

    \subsection{Implementation}

We implemented \nemo{} on top of Block-STM~\cite{Block-STM} in Rust using around 3,500 lines of code~\footnotemark{}.
We use simulated execution instead of actually executing transactions, as the improvements made by \nemo{} do not target the execution of transactions inside the virtual machine. We simulate execution by making a thread sleep for a certain amount of time and record the read/write sets of a transaction. 
We randomly sampled this simulated execution duration given in milliseconds following a $LogNormal(2.0, 0.5)$ distribution in order to have some variation. This gave us an expected execution time per transaction of around 8.4 ms, a median of around 7.4 ms and a 90th percentile of around 14.0 ms.

\footnotetext{https://anonymous.4open.science/r/nemo-DFC2}

\begin{figure*}[h]
  \centering
  \begin{minipage}[b]{0.49\textwidth}
    \centering
    \includegraphics[width=\columnwidth]{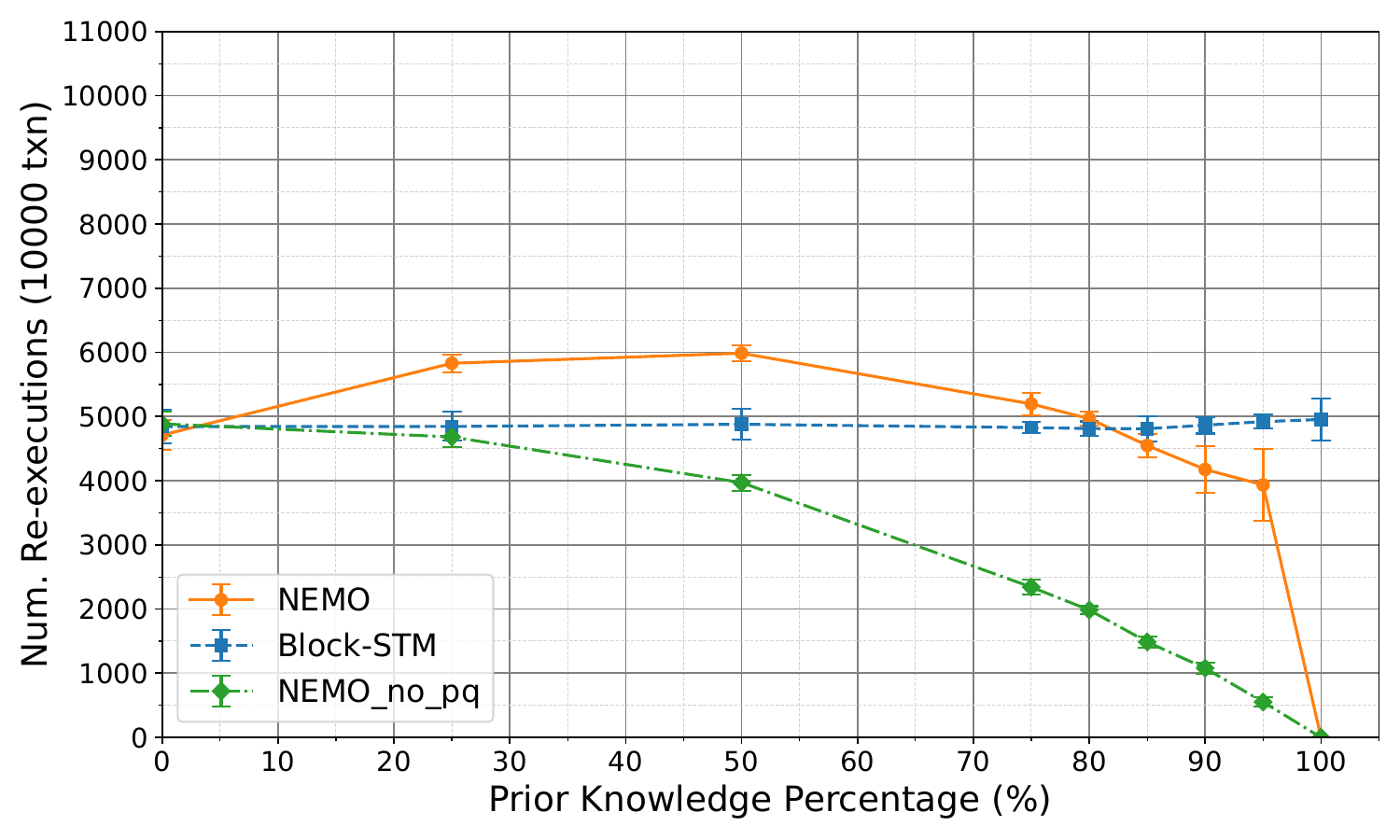}
    \caption*{(a) 8 Workers}
  \end{minipage}
  \hfill
  \begin{minipage}[b]{0.49\textwidth}
    \centering
    \includegraphics[width=\columnwidth]{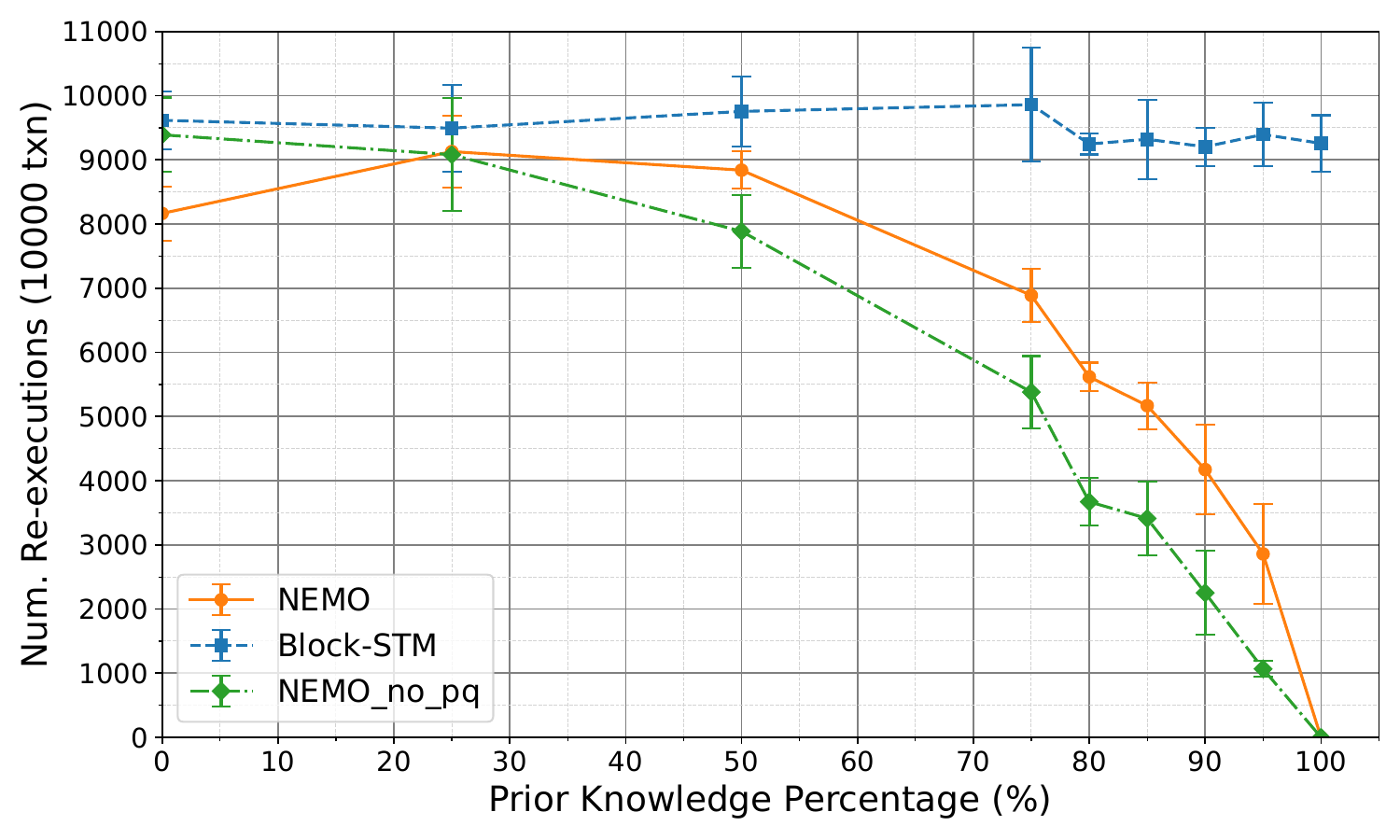}
    \caption*{(b) 16 Workers}
  \end{minipage}
  \caption{Number of re-executions depending on assumed proportion of prior knowledge}
  \label{paper:fig:high-num-reexecutions}
\end{figure*}

To accurately simulate the execution of transactions, the exhaustive read/write sets of objects that a transaction can use is determined when generating the workload. Each object in these sets then has a 90\% chance of actually being used by the transaction during execution. We then vary a parameter that controls the likelihood of an object that is used to be apart of the prior knowledge read/write sets. This parameter is a value between 0 and 100, where 0 means that there is no prior knowledge available and 100 means that there is complete prior knowledge of all objects used. This allows us to simulate having partial knowledge of a transaction's read/write sets ahead of time.

    \subsection{Workloads}

To evaluate \nemo{} under high contention, we consider a synthetic scenario with 50 shared objects where each transaction accesses a number of objects that is sampled from $LogNormal(0.5, 0.5)$. This gives an expected value of around $1.87$, a median of around $1.65$ and a 90th percentile of around $3.13$. For each object accessed we then sample from $Zipf(50, 2.0)$ which is a skewed distribution to simulate hot objects causing contention between transactions. Using this method, the hottest object has around a 63\% chance of being sampled, the second hottest around a 15\% chance and the third around a 7\% chance of being sampled. Each object access by a transaction then has a 35\% chance of being only a read, a 42.25\% chance of being a read + write, and a 22.75\% chance of being only a write. These setting simulate the highly skewed data access that are a defining feature of blockchain workloads \cite{ParallelEVM}.

The main metric used to evaluate performance is the duration taken to execute a block. Taking the block size and dividing it by the duration then gives us the \textit{throughput} measured in transactions per second (TPS). \nemo{} and Block-STM both use a lazy commit except that \nemo{} also has the \textit{greedy commit rule}, which means that if \nemo{} is able to achieve a better throughput it would also mean it achieved a better latency than Block-STM. We also measured the number of re-executions to track how much redundant work has been done, because we designed \nemo{} to minimize this metric in hopes that it would lead to an increase in performance.

All experiments were run using a supercomputer 
with 2GB of memory per CPU. All code was ran using release mode. For every data point collected, we plot the mean and standard deviation of 5 runs. 
We varied the number of workers that can execute tasks as well as the percentage of prior knowledge available to see what effect it has on the performance of the system. In our experiments we compare \nemo{} to sequential execution, Block-STM, and a PCC baseline. The PCC baseline looks to simulate Sui Lutris~\cite{SuiLutris} execution by using the exhaustive read/write sets to prevent potentially conflicting transactions from executing in parallel. Transactions are scheduled based on when they became ready for execution and are assigned to the first available worker.
We consider the full version of \nemo{}, and \nemonopq{}, its version without the priority-based scheduling that uses a priority queue.

\begin{figure*}[h]
  \centering
  \begin{minipage}[b]{0.49\textwidth}
    \centering
    \includegraphics[width=\columnwidth]{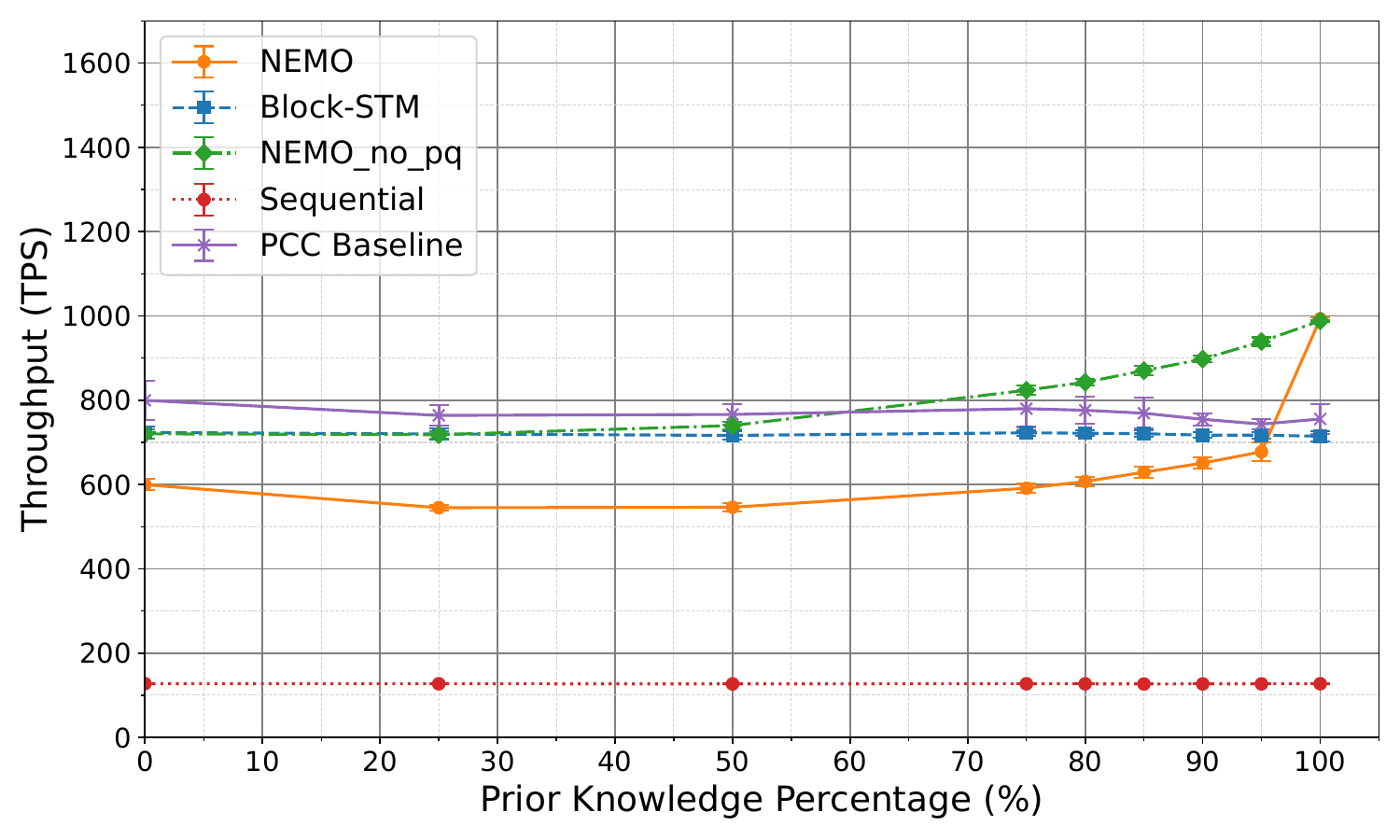}
    \caption*{(a) 8 Workers}
  \end{minipage}
  \hfill
  \begin{minipage}[b]{0.49\textwidth}
    \centering
    \includegraphics[width=\columnwidth]{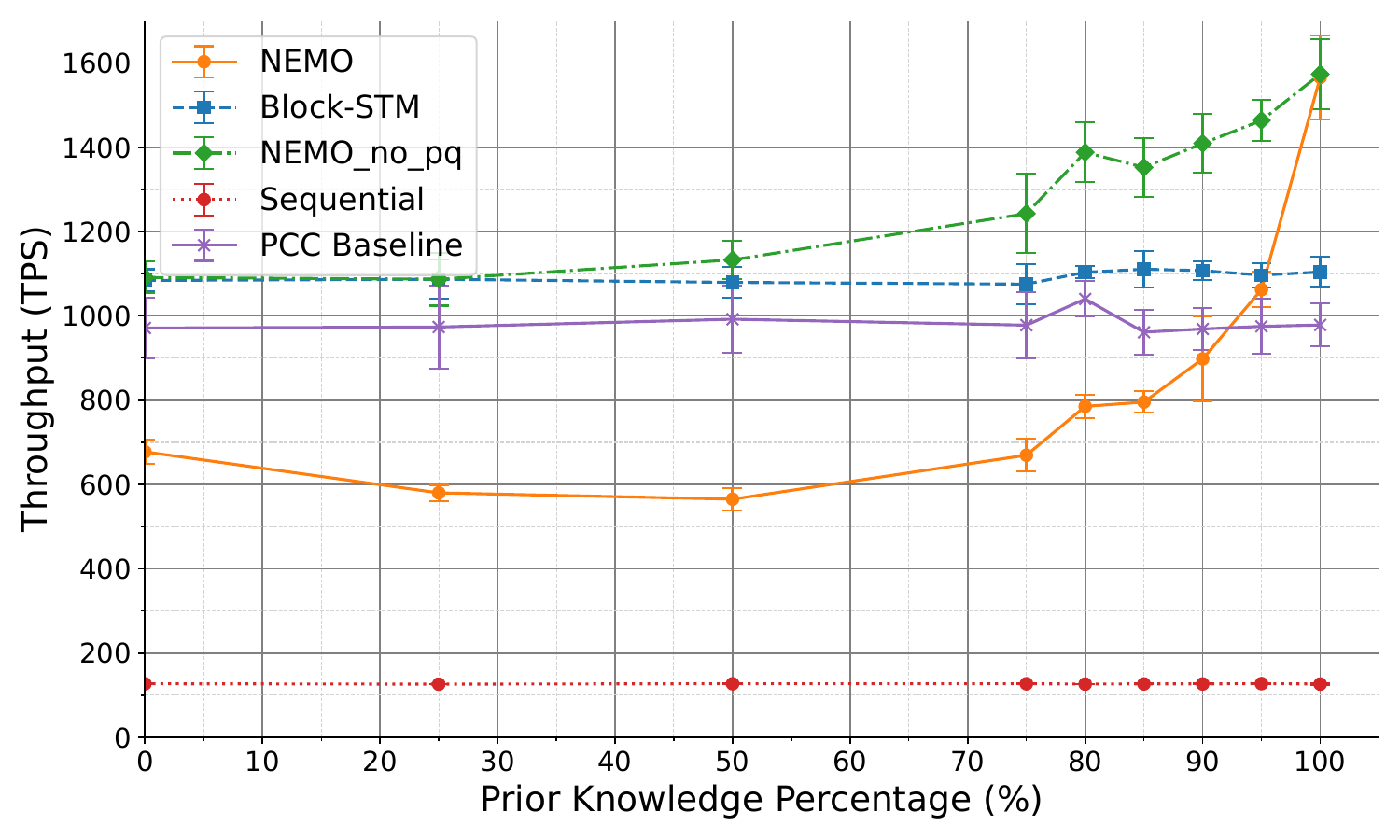}
    \caption*{(b) 16 Workers}
  \end{minipage}
  \caption{Execution throughput depending on assumed proportion of prior knowledge}
  \label{paper:fig:high-throughput}
\end{figure*}

    \subsection{Number of re-executions}

Figure~\ref{paper:fig:high-num-reexecutions} shows the number of re-executions with 8 and 16 workers for Block-STM and \nemo{}. Other  baselines, i.e., sequential execution and PCC baseline, never re-execute transactions and are therefore not shown. More prior knowledge effectively reduces the number of re-executions for \nemo{}, especially when priority scheduling is disabled, down to 0 with full knowledge.
\nemo{} with 16 workers, and \nemonopq{} with 8 or 16 workers, always outperform Block-STM. With 8 workers, \nemo{} requires at least 85\% of prior knowledge to outperform Block-STM. Using more worker threads improves \nemo{}'s performance compared to Block-STM. 
There are more re-executions when priority scheduling is enabled. Whilst this may be surprising at first it makes sense, as the priority scheduling is based on the number of known dependencies. This means that a transaction ordered later that is known to block another one will be scheduled before another transaction that is ordered earlier and for which no dependencies are known. This is the intended behavior, but in a high contention setting with limited prior knowledge it often means that the earlier transaction also blocks other transactions that \nemo{} does not know about. 
\nemonopq{} is able to halve the number of re-executions when given 75\% prior knowledge for both 8 and 16 workers.
Increasing the number of workers increases the number of re-executions for all the protocols.

    \subsection{Throughput}
    
Figure~\ref{paper:fig:high-throughput} shows the throughput of the protocols. We can see that increasing prior knowledge leads to increased throughput and that even partial knowledge can lead to a significant performance gain. For 16 workers we can see that \nemo{} already significantly outperforms Block-SMT at 75\% prior knowledge with a throughput of 1243 TPS compared to 1076 TPS for Block-STM which represents a 15.5\% improvement in throughput, and even more at 90\% with a throughput of 1409 TPS compared to 1108 TPS for Block-STM which is a 27.1\% improvement. The results also show that priority scheduling is only viable when a very high percentage of prior knowledge is available and explains why Chiron~\cite{Chiron} was only used for straggler execution. When full prior knowledge is available \nemo{} achieves 993 TPS for 8 workers as opposed to 715 TPS for Block-STM which is a 38.9\% improvement and achieves 1574 TPS for 16 workers as opposed to 1105 TPS for Block-STM which is a 42.4\% improvement. For comparison, Chiron only improves over Block-STM by 30\%. When comparing to the PCC baseline that achieved 979 TPS, this represents a 60.8\% improvement in throughput.

\section{Discussion and Future Work}
Our evaluation demonstrates that \texttt{NEMO} effectively advances the state of the art in blockchain execution for high-contention workloads. By leveraging partial prior knowledge about transactions’ read/write sets, \texttt{NEMO} significantly reduces the number of failed validations and re-executions compared to the optimistic baseline Block-STM. This reduction in redundant work directly translates into improved throughput, especially as contention levels increase. In high-contention scenarios, \texttt{NEMO} achieves up to 60.8\% higher throughput compared to the PCC baseline and 50.1\% higher throughput relative to Block-STM with 90\% prior knowledge, underscoring the protocol’s effectiveness in managing conflicts while maintaining optimistic concurrency.

The evaluation also highlights that the benefits of priority scheduling depend strongly on the completeness of prior knowledge. While priority scheduling can deliver the best performance when full prior knowledge is available, it performs poorly under partial knowledge. This suggests that \texttt{NEMO} can benefit from adaptive scheduling strategies that select appropriate policies based on the amount and quality of prior knowledge, enabling more robust performance across varying workload conditions.

Additionally, we observed that increasing the number of workers generally improves throughput, but this benefit plateaus due to increased contention and the resulting overhead of managing conflicts. This highlights the importance of carefully tuning system parameters such as worker count to balance parallelism gains against contention costs.

Several avenues remain open for future work. Evaluating \texttt{NEMO} on real-world blockchain workloads would provide valuable insights into its practical applicability and performance under production conditions. Furthermore, developing static analysis techniques to automatically extract partial prior knowledge from smart contract logic would reduce manual effort and improve the protocol’s adaptability. Finally, extending \texttt{NEMO} beyond shared-memory environments, for example by designing sharded or distributed execution strategies tailored for high-contention workloads, presents an exciting opportunity to further scale blockchain throughput.

\section{Conclusion}

This work presented \nemo{}, a novel blockchain execution engine that integrates the object data model with optimistic concurrency control (OCC) to achieve high throughput under highly contended workloads. By adapting and enhancing the Block-STM protocol to the object data model, \nemo{} introduces several key innovations, including a greedy commit rule for independent transactions, improved dependency tracking, partial prior knowledge utilization, and intelligent priority scheduling.
Our evaluation demonstrates that even partial prior knowledge of transaction object access patterns significantly reduces redundant re-executions and improves throughput. \nemo{} consistently outperforms existing baselines under high contention, with up to 42\% improvement in throughput compared to Block-STM and 61\% compared to the PCC baseline. 

\printbibliography

@inproceedings{Block-STM,
author = {Gelashvili, Rati and Spiegelman, Alexander and Xiang, Zhuolun and Danezis, George and Li, Zekun and Malkhi, Dahlia and Xia, Yu and Zhou, Runtian},
title = {Block-STM: Scaling Blockchain Execution by Turning Ordering Curse to a Performance Blessing},
year = {2023},
booktitle = {PPoPP}
}

@inproceedings{PEEP,
author = {Chen, Zhihao and Qi, Xiaodong and Du, Xiaofan and Zhang, Zhao and Jin, Cheqing},
title = {PEEP: A Parallel Execution Engine for Permissioned Blockchain Systems},
year = {2021},
booktitle = {DASFAA}
}

@misc{Pilotfish,
      title={Pilotfish: Distributed Transaction Execution for Lazy Blockchains}, 
      author={Quentin Kniep and Lefteris Kokoris-Kogias and Alberto Sonnino and Igor Zablotchi and Nuda Zhang},
      year={2024},
      eprint={2401.16292},
      archivePrefix={arXiv},
      url={https://arxiv.org/abs/2401.16292}, 
}

@inproceedings{OptME,
    author = {Ryu, Donghyeon and Park, Chanik},
    title = {Toward High-Performance Blockchain System by Blurring the Line between Ordering and Execution},
    year = {2024},
    booktitle = {SC}
}

@mastersthesis{Block-X,
    author = {Rizwan Shahid},
    title = {Parallel Transaction Execution in Public Blockchain Systems},
    school = {University of Waterloo},
    year = {2024}
}

@inproceedings{Narwhal,
author = {Danezis, George and Kokoris-Kogias, Lefteris and Sonnino, Alberto and Spiegelman, Alexander},
title = {Narwhal and Tusk: a DAG-based mempool and efficient BFT consensus},
year = {2022},
booktitle = {EuroSys}
}

@INPROCEEDINGS {ParBlockchain,
author = { Amiri, Mohammad Javad and Agrawal, Divyakant and El Abbadi, Amr },
booktitle = {ICDCS},
title = {{ ParBlockchain: Leveraging Transaction Parallelism in Permissioned Blockchain Systems }},
year = {2019}
}

@misc{Mysticeti,
      title={Mysticeti: Reaching the Limits of Latency with Uncertified DAGs}, 
      author={Kushal Babel and Andrey Chursin and George Danezis and Anastasios Kichidis and Lefteris Kokoris-Kogias and Arun Koshy and Alberto Sonnino and Mingwei Tian},
      year={2024},
      eprint={2310.14821},
      archivePrefix={arXiv},
      url={https://arxiv.org/abs/2310.14821}, 
}

@misc{Arete,
      title={Optimal Sharding for Scalable Blockchains with Deconstructed SMR}, 
      author={Jianting Zhang and Zhongtang Luo and Raghavendra Ramesh and Aniket Kate},
      year={2024},
      eprint={2406.08252},
      archivePrefix={arXiv},
      url={https://arxiv.org/abs/2406.08252}, 
}

@misc{SuiLutris,
      title={Sui Lutris: A Blockchain Combining Broadcast and Consensus}, 
      author={Sam Blackshear and Andrey Chursin and George Danezis and Anastasios Kichidis and Lefteris Kokoris-Kogias and Xun Li and Mark Logan and Ashok Menon and Todd Nowacki and Alberto Sonnino and Brandon Williams and Lu Zhang},
      year={2024},
      eprint={2310.18042},
      archivePrefix={arXiv},
      url={https://arxiv.org/abs/2310.18042}, 
}

@misc{RapidLane,
      title={Deferred Objects to Enhance Smart Contract Programming with Optimistic Parallel Execution}, 
      author={George Mitenkov and Igor Kabiljo and Zekun Li and Alexander Spiegelman and Satyanarayana Vusirikala and Zhuolun Xiang and Aleksandar Zlateski and Nuno P. Lopes and Rati Gelashvili},
      year={2024},
      eprint={2405.06117},
      archivePrefix={arXiv},
      url={https://arxiv.org/abs/2405.06117}, 
}

@misc{SuiWhitepaper,
  title={The Sui Smart Contracts Platform},
  author={The MystenLabs Team},
  year={2022},
  note={\url{https://docs.sui.io/paper/sui.pdf}},
}

@inproceedings{MoveVm,
  title={Move: A Language With Programmable Resources},
  author={Sam Blackshear and Evan Cheng and David L. Dill and Victor Gao and Ben Maurer and Todd Nowacki and Alistair Pott and Shaz Qadeer and Dario Rain and Stephane Russi and Tim Sezer and Runtian Zakian and Zhou},
  year={2019},
  url={https://api.semanticscholar.org/CorpusID:201681125}
}

@misc{Shardines,
  title = {Shardines: Aptos’ Sharded Execution Engine Blazes to 1M TPS},
  author = {Manu Dhundi and Sital Kedia and Igor Kabiljo and Zhoujun Ma and Jan Olkowski},
  year = {2025},
  month = {2},
  url = {https://aptoslabs.medium.com/shardines-aptos-sharded-execution-engine-blazes-to-1m-tps-71c5f9b8bf60}
}

@inproceedings{Bullshark,
author = {Spiegelman, Alexander and Giridharan, Neil and Sonnino, Alberto and Kokoris-Kogias, Lefteris},
title = {Bullshark: DAG BFT Protocols Made Practical},
year = {2022},
booktitle = {CCS}
}

@article{HotStuff,
  author       = {Ittai Abraham and
                  Guy Gueta and
                  Dahlia Malkhi},
  title        = {Hot-Stuff the Linear, Optimal-Resilience, One-Message {BFT} Devil},
  journal      = {CoRR},
  volume       = {abs/1803.05069},
  year         = {2018},
  url          = {http://arxiv.org/abs/1803.05069},
  eprinttype    = {arXiv},
  eprint       = {1803.05069},
  bibsource    = {dblp computer science bibliography, https://dblp.org}
}

@misc{AptosWhitepaper,
  title={The Aptos Blockchain: Safe, Scalable, and Upgradeable Web3 Infrastructure},
  year={2022},
  note={\url{https://aptosfoundation.org/whitepaper/aptos-whitepaper_en.pdf}},
}

@article{Concerto,
  title={Concerto: Transaction-Parallel EVM},
  author={Su, Stephen and Hu, Jeffrey and Govil, Yashodhar and Kohli, Sumer},
  journal={Stanford University},
  year={2024},
  url={https://www.scs.stanford.edu/24sp-cs244b/projects/Concerto_Transaction_Parallel_EVM.pdf}
}

@inproceedings{Forerunner,
author = {Chen, Yang and Guo, Zhongxin and Li, Runhuai and Chen, Shuo and Zhou, Lidong and Zhou, Yajin and Zhang, Xian},
title = {Forerunner: Constraint-based Speculative Transaction Execution for Ethereum},
year = {2021},
booktitle = {SOSP}
}

@misc{MBPS,
      title={Unleashing Multicore Strength for Efficient Execution of Transactions}, 
      author={Ankit Ravish and Akshay Tejwani and Piduguralla Manaswini and Sathya Peri},
      year={2024},
      eprint={2410.22460},
      archivePrefix={arXiv},
      url={https://arxiv.org/abs/2410.22460}, 
}

@article{Dipetrans,
author = {Baheti, Shrey and Parwat Singh, Anjana and Peri, Sathya and Simmhan, Yogesh},
year = {2022},
month = {01},
pages = {},
title = {DiPETrans: A framework for distributed parallel execution of transactions of blocks in blockchains},
volume = {34},
journal = {Concurrency and Computation: Practice and Experience},
doi = {10.1002/cpe.6804}
}

@MASTERSTHESIS{MeteringTheMeter,
	copyright = {In Copyright - Non-Commercial Use Permitted},
	year = {2023},
        month = {10},
	type = {Master Thesis},
	author = {Mitenkov, George},
	size = {99 p.},
	publisher = {ETH Zurich},
	DOI = {10.3929/ethz-b-000638680},
	title = {Metering the Meter, or How to Efficiently and Deterministically Charge the Execution of Smart Contracts},
	school = {ETH Zurich}
}

@article{BitcoinWhitepaper,
  author = {Nakamoto, Satoshi},
  month = may,
  title = {Bitcoin: A Peer-to-Peer Electronic Cash System},
  url = {http://www.bitcoin.org/bitcoin.pdf},
  year = 2009
}

@INPROCEEDINGS{DMVCC,
  author={Qi, Xiaodong and Jiao, Jiao and Li, Yi},
  booktitle={ICDCS}, 
  title={Smart Contract Parallel Execution with Fine-Grained State Accesses}, 
  year={2023},
  doi={10.1109/ICDCS57875.2023.00068}
}

@article{Dodo,
title = {Dodo: A scalable optimistic deterministic concurrency control protocol},
journal = {Future Generation Computer Systems},
volume = {159},
pages = {15-26},
year = {2024},
issn = {0167-739X},
doi = {https://doi.org/10.1016/j.future.2024.05.004},
url = {https://www.sciencedirect.com/science/article/pii/S0167739X24002139},
author = {Xinyuan Wang and Yun Peng and Hejiao Huang and Xingchen Li}
}

@article{DOCC,
author = {Dong, Zhi-Yuan and Tang, Chu-Zhe and Wang, Jia-Chen and Wang, Zhao-Guo and Chen, Hai-Bo and Zang, Bin-Yu},
title = {Optimistic Transaction Processing in Deterministic Database},
year = {2020},
publisher = {Springer-Verlag},
address = {Berlin, Heidelberg},
volume = {35},
number = {2},
issn = {1000-9000},
url = {https://doi.org/10.1007/s11390-020-9700-5},
doi = {10.1007/s11390-020-9700-5},
journal = {J. Comput. Sci. Technol.},
month = mar,
pages = {382–394},
numpages = {13}
}

@misc{Anthemius,
      title={Anthemius: Efficient \& Modular Block Assembly for Concurrent Execution}, 
      author={Ray Neiheiser and Eleftherios Kokoris-Kogias},
      year={2025},
      eprint={2502.10074},
      archivePrefix={arXiv},
      url={https://arxiv.org/abs/2502.10074}, 
}

@misc{Chiron,
      title={CHIRON: Accelerating Node Synchronization without Security Trade-offs in Distributed Ledgers}, 
      author={Ray Neiheiser and Arman Babaei and Giannis Alexopoulos and Marios Kogias and Eleftherios Kokoris Kogias},
      year={2024},
      eprint={2401.14278},
      archivePrefix={arXiv},
      url={https://arxiv.org/abs/2401.14278}, 
}

@inproceedings{Ethereum,
  title={Ethereum: a secure decentralised generalised transaction ledger},
  author={},
  year={2019},
  url={https://api.semanticscholar.org/CorpusID:261284440}
}

@INPROCEEDINGS {Batch-Schedule-Execute,
author = { Hay, Yaron and Friedman, Roy },
booktitle = {SRDS},
title = {{ Batch-Schedule-Execute: On Optimizing Concurrent Deterministic Scheduling for Blockchains }},
year = {2024}
}

@article{Gria,
  title={Gria: an efficient deterministic concurrency control protocol},
  author={Xinyuan Wang and Yun Peng and Hejiao Huang},
  journal={Frontiers of Computer Science},
  year={2023},
  volume={18},
  pages={1-13},
  url={https://api.semanticscholar.org/CorpusID:266344805}
}

@misc{Thunderbolt,
      title={Thunderbolt: Concurrent Smart Contract Execution with Nonblocking Reconfiguration for Sharded DAGs}, 
      author={Junchao Chen and Alberto Sonnino and Lefteris Kokoris-Kogias and Mohammad Sadoghi},
      year={2025},
      eprint={2407.09409},
      archivePrefix={arXiv},
      url={https://arxiv.org/abs/2407.09409}, 
}

@inproceedings{LazyBlockchainClient,
author = {Tas, Ertem Nusret and Tse, David and Yang, Lei and Zindros, Dionysis},
title = {Light Clients for Lazy Blockchains},
year = {2025},
booktitle = {FC}
}

@article{AllAboutObjects,
  title     = {All About Objects},
  journal   = {Sui Foundation},
  year      = {2023},
  url       = {https://blog.sui.io/all-about-objects/}
}

@inproceedings{Chartalist,
 author = {Shamsi, Kiarash and Victor, Friedhelm and Kantarcioglu, Murat and Gel, Yulia and Akcora, Cuneyt G},
 booktitle = {NeurIPS},
 title = {Chartalist: Labeled Graph Datasets for UTXO and Account-based Blockchains},
 year = {2022}
}

@misc{UTXOvsAccount,
      title={UTXO in Digital Currencies: Account-based or Token-based? Or Both?}, 
      author={Aldar C-F. Chan},
      year={2021},
      eprint={2109.09294},
      archivePrefix={arXiv},
      url={https://arxiv.org/abs/2109.09294}, 
}

@misc{Stingray,
      title={Stingray: Fast Concurrent Transactions Without Consensus}, 
      author={Srivatsan Sridhar and Alberto Sonnino and Lefteris Kokoris-Kogias},
      year={2025},
      eprint={2501.06531},
      archivePrefix={arXiv},
      url={https://arxiv.org/abs/2501.06531}, 
}

@inproceedings{ParallelEVM,
author = {Lin, Haoran and Feng, Hang and Zhou, Yajin and Wu, Lei},
title = {ParallelEVM: Operation-Level Concurrent Transaction Execution for EVM-Compatible Blockchains},
year = {2025},
isbn = {9798400711961},
publisher = {Association for Computing Machinery},
address = {New York, NY, USA},
url = {https://doi.org/10.1145/3689031.3696063},
doi = {10.1145/3689031.3696063},
booktitle = {Proceedings of the Twentieth European Conference on Computer Systems},
pages = {211–225},
numpages = {15},
location = {Rotterdam, Netherlands},
series = {EuroSys '25}
}

@inproceedings{OCCDA,
author = {Garamv\"{o}lgyi, P\'{e}ter and Liu, Yuxi and Zhou, Dong and Long, Fan and Wu, Ming},
title = {Utilizing parallelism in smart contracts on decentralized blockchains by taming application-inherent conflicts},
year = {2022},
booktitle = {ICSE}
}

@online{iota_rebased_2024,
  author       = {{IOTA Foundation}},
  title        = {IOTA Rebased: Fast Forward},
  year         = 2024,
  url          = {https://blog.iota.org/iota-rebased-fast-forward},
  note         = {Accessed: 2025-06-15}
}

\end{document}